\documentclass[12pt,preprint]{aastex}

\newcommand{\ba}{\begin{eqnarray}}
\newcommand{\ea}{\end{eqnarray}}
\newcommand{\be}{\begin{equation}}
\newcommand{\ee}{\end{equation}}

\begin{document}

\title{Time Delay of Cascade Radiation for TeV Blazars and 
the Measurement of the Intergalactic Magnetic Field}
 
\author{Charles D. Dermer\altaffilmark{1}, Massimo Cavadini\altaffilmark{2}, Soebur Razzaque\altaffilmark{1,3},
Justin D.\ Finke\altaffilmark{1}, James Chiang\altaffilmark{4} \& Benoit Lott\altaffilmark{5,6}}
 
\altaffiltext{1}{Space Science Division, U.S. Naval
  Research Laboratory, Washington, DC 20375, USA. e-mail:
  charles.dermer@nrl.navy.mil }
\altaffiltext{2}{Dipartimento di Fisica e Matematica, Universit\'a dell'Insubria, via Valleggio 11, 22100, Como, Italy }
\altaffiltext{3}{NRL/NRC Resident Research Associate }
\altaffiltext{4}{W.\ W.\ Hansen Experimental Physics Laboratory, Kavli Institute for Particle Astrophysics and Cosmology, Department of Physics and SLAC National Accelerator
Laboratory, Stanford University, Stanford, CA 94305, USA}
\altaffiltext{5}{CNRS/IN2P3, Centre d' \'Etudes Nucl\'eaires Bordeaux Gradignan, UMR 5797, Gradignan, 33175, France}
\altaffiltext{6}{40 Universit\'e de Bordeaux, Centre d' \'Etudes Nucl\'eaires Bordeaux Gradignan, UMR 5797, Gradignan, 33175, France}

\begin{abstract} 
Recent claims that the strength $B_{\rm IGMF}$ of the intergalactic magnetic field (IGMF) 
is $\gtrsim 10^{-15}$ G are based on upper limits to the expected cascade flux in the GeV band produced by 
blazar TeV photons absorbed by the extragalactic background light.  This limit depends 
on an assumption that the mean blazar TeV flux remains constant on timescales
$\gtrsim 2 (B_{\rm IGMF}/10^{-18}{\rm~G})^2/(E/{\rm 10~GeV})^2$ yr for an IGMF coherence length $\approx 1$ Mpc, where $E$ is the measured photon energy.
{ Restricting  TeV activity of 1ES 0229+200 to $\approx 3$ -- 4 years during which the source has been observed leads to a more robust lower limit of $B_{IGMF} \gtrsim 10^{-18}$ G, which can be larger by an order of magnitude if the intrinsic source flux above $\approx 5$ -- 10 TeV  from 1ES 0229+200 is strong.}
\end{abstract}

\keywords{gamma rays: theory---radiation mechanisms: nonthermal}

\section{Introduction}

The measurement of the intergalactic magnetic field (IGMF) gives information about processes operating in 
the early universe that are imprinted on the large scale structure of the universe \citep[see][]{ns09}. 
Faraday rotation measurements of the radio emission of quasars,
patterns in the arrival directions of UHECRs towards the supergalactic plane 
and Cen A,
and theoretical arguments from COBE data  \citep{bfs97} indicate that $B_{\rm IGMF}\ll 10^{-8}$ G, 
but no direct measurements or lower limits of the IGMF in the voids have been firmly established. Gamma-ray astronomy provides a method to measure the IGMF through magnetic field-induced delays  \citep{pla95}, 
pair halos from sources of TeV photons directed away from our line of sight \citep{acv94}, and  halos around \citep{ens09} and cascade-radiation spectra of \citep{dai02,adg07,mur08,nv10,tav10a,tav10b} point sources of high-energy $\gamma$ rays. 

In the latter approach, TeV photons from cosmic $\gamma$-ray sources interact with photons of the 
extragalactic background light (EBL) to produce e$^-$e$^+$ pairs. If the pairs are not significantly deflected
by the IGMF, 
cosmic microwave background (CMB) photons, which dominate the radiation energy density
in intergalactic space, are Compton-scattered in 
the original direction of the pairs. Deflection out
of the TeV beam depends on the jet opening angle $\theta_{\rm j}$.
By comparing blazar TeV fluxes with upper limits on the GeV radiation flux measurements with the Fermi 
Gamma ray Space Telescope, several claims (see Table 1) have been made that lower limits on the IGMF have been measured
\citep{nv10,tav10a,tav10b,dol10}. These limits, which depend on the assumed opening angle $\theta_{\rm j}$  of the TeV photon source, are summarized in Table 1 for the TeV blazar 1ES 0229+200 at redshift $z = 0.1396$. Under the assumption of persistent TeV emission over long time scales,
these studies find that   $B_{\rm IGMF} \approx 10^{-15}$ G for magnetic coherence (or correlation) lengths $\lambda_{coh} \sim 1$ Mpc
when  $\theta_{\rm j} \cong 0.1$.

Implicit in all these studies is that the TeV blazars used to infer the IGMF emit constant flux over a long period of time. Because blazars are highly variable, a { more defensible limit is obtained (lacking other ways to infer a blazar's TeV activity lifetime) by assuming that the TeV 
emission is emitted only 
over the past few years during which it has been monitored. A simple semi-analytic approach is used to derive minimum values, for comparison with 
numerical models \citep[][]{dol10,tvn11}. }

\section{Time Delay and Deflection Angle of Emission from First Generation Pairs}

Consider a source and observer separated by distance $d$, as shown in Figure 1. 
Photons with dimensionless energy $\epsilon_1 = h\nu_1/m_ec^2 \cong 2\times 10^6 E_1$(TeV) emitted at angle $\theta_1$ to the line of sight between the source and observer, travel a mean distance $\lambda_{\gamma\gamma}= \lambda_{\gamma\gamma}(\epsilon_1,z)$ before converting into an electron-positron pair via $\gamma\gamma$ absorption with photons of the EBL. The pairs cool by scattering CMB photons to $E_{\rm GeV}$ GeV energies, which  are detected at an angle $\theta$ to the line of sight to the source when the secondary electrons and positrons (hereafter referred to as electrons) are deflected by an angle $\theta_{\rm dfl}$. { The GeV emission, to be detected, must be within the energy-dependent Fermi LAT point-spread-function angle $\theta_{psf}(E_{\rm GeV})$.
The system is treated in the low redshift limit}  \citep[cf.][]{ns09}.

\begin{figure}[t]
% \vspace*{-2.0 cm}
\begin{center}
 \includegraphics[width=3.4in]{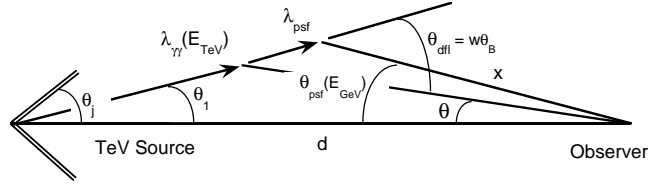} 
% \vspace*{-1.0 cm}
 \caption{Sketch of the geometry of the process. 
A photon with energy $E_{\rm TeV}$ TeV, 
emitted at angle $\theta_1\leq \theta_{\rm j}$ to the line of sight, 
interacts with an EBL photon to create an electron-positron pair
with Lorentz factor $\gamma = 10^6 \gamma_6$. The lepton is deflected 
through angle $\theta_{\rm dfl}$ and scatters a CMB photon to 
energy $E_{\rm GeV}$ GeV, { which is observed as a source photon by the Fermi LAT 
if it is detected at an angle $\theta < \theta_{psf}(E_{\rm GeV})$ to the source. 
The underlying simplifying kinematic relation in the semi-analytic model is
$\gamma_6 \approx E_{\rm TeV} \approx \sqrt{{\rm E}_{\rm GeV}}$}.
}
\label{fig1}
\end{center}
\end{figure}

The time delay $\Delta t$ between the reception of photons directed towards the observer and those formed
by the process described above is given by
$$c\Delta t = \lambda_{\gamma\gamma} + x - d = \lambda_{\gamma\gamma} + {d\sin(\theta_{\rm dfl}-\theta)\over \sin\theta_{\rm dfl}}
 - d = $$
\begin{equation}
\lambda_{\gamma\gamma}(1-\cos\theta_{\rm dfl}) - d(1-\cos\theta)\;,
\label{cDeltat}
\end{equation} 
noting that $x = d\sin\theta_1/\sin\theta_{\rm dfl}$ and $\lambda_{\gamma\gamma} = d\sin \theta/\sin \theta_{\rm dfl}$.
In the limit of small observing and deflection angles, equation (\ref{cDeltat}) implies 
\begin{equation}
\Delta t \cong {\lambda_{\gamma\gamma}\over 2c}\;\theta_{\rm dfl}^2\;,
\label{Deltat}
\end{equation}
provided that the photon is detected at an angle
\begin{equation}
\theta =
{\lambda_{\gamma\gamma}({\rm E}_{\rm TeV})\theta_{dfl}({\rm E}_{\rm GeV})\over d}
<  \theta_{psf}({\rm E}_{\rm GeV}) \;
\label{theta}
\end{equation}
to the source.
%, where the energy-dependent 
%Fermi LAT point spread function is denoted by $\theta_{psf}({\rm E}_{\rm GeV})$. 
Note that the deflection angle depends on either
the primary photon energy ${\rm E}_{\rm TeV}$ or Compton-scattered
photon energy ${\rm E}_{\rm GeV}$, since they 
are related by ${\rm E}_{\rm GeV} \approx {\rm E}_{\rm TeV}^2$, as 
we now show.

\begin{table}% use packages: array
\caption{Derived Limits on $B_{\rm IGMF}$ for the source 1ES 0229+200}
\begin{center}
\begin{tabular}{|l|l|l|}
\hline
1ES 0229+200 & $\theta_{\rm j}$ (rad)& $B_{\rm IGMF}$(G) \\ 
\hline
\citet{nv10} & $\pi$ & $\gtrsim 3\times 10^{-16}$ 
  \\
%\hline
\citet{tav10a} & 0.1 & $\gtrsim  5\times 10^{-15}$ \\
\citet{tav10b} & 0.03 & $\gtrsim  2\times 10^{-15}$ \\
%\hline
\cite{dol10} & 0.1  & $\gtrsim  5\times 10^{-15}$  \\ 
\hline
\end{tabular}
\end{center}
\label{car}
\end{table}

The average CMB photon energy at low redshift is $\epsilon_{0}\approx 1.24\times 10^{-9}$ in $m_ec^2$ units, so that  mean Thomson-scattered photon energy is $\epsilon_{\rm T}\approx (4/3)\epsilon_{0}\gamma^2$, where $\gamma \cong E_{\rm TeV}/(2 m_ec^2)$ implies
$\gamma_6 \cong 0.98 E_{\rm TeV}$. Thus, an electron with Lorentz factor $\gamma$ scatters CMB radiation to photon energy $E$ when $\gamma_6 \cong E_{\rm TeV} \cong 1.1\sqrt{E_{\rm GeV}}$. The characteristic length scale for energy losses due to Thomson scattering is $\lambda_{\rm T} = 3m_ec^2/4\sigma_{\rm T} u_{\rm CMB}\gamma = 
(0.75/\gamma_6)$ Mpc, where $u_{\rm CMB} \cong 4\times 10^{-13}$ erg cm$^{-3}$ is the CMB energy density at low redshifts. While losing energy, the electron is deflected by an angle $\theta_{\rm B}\cong \lambda_{\rm T}/r_{\rm L}$ in a uniform magnetic field of strength $B_{\rm IGMF} = 10^{-15} B_{-15}$ G oriented perpendicular to the direction of motion of the electron, where the Larmor radius  $r_{\rm L} = m_ec^2\gamma/eB \cong 0.55 (\gamma_6/B_{-15})$ Mpc. Thus the deflection angle for an electron losing energy by scattering CMB photons to energy $E$ in a uniform field is $\theta_B = \lambda_{\rm T}/r_{\rm L}\cong 1.1B_{-15}/E_{\rm GeV}$. Introducing a coherence length $\lambda_{coh}$ that characterizes the typical distance over which the magnetic field direction changes by $\approx \pi/2$, then the deflection angle
\begin{equation}
\theta_{\rm dfl} \equiv w\theta_{\rm B}\;,\;{\rm with}\;\;w = \cases{~~~1 & if $\lambda_{\rm T}< \lambda_{coh}$\cr
	\sqrt{\lambda_{coh}\over \lambda_{\rm T}},& if $\lambda_{\rm T}> \lambda_{coh}$.\cr} \;.
\label{thetadlf}
\end{equation}

For 1ES 0229+200, TeV radiation has been detected to energies  $E\lesssim 12$ TeV \citep{aha07}, with an $\approx 15$\% error 
in the energy measurement. 
 An uncertainty in the analytic treatment 
is that the mean free path $\lambda_{\gamma\gamma}(E_{\rm TeV}$) varies by a factor of $\approx 2$ between $z \rightarrow 0$ 
and $z =0.14$, and between different EBL models. 
For instance, 
the EBL model of \citet{frd10}  gives $\lambda_{\gamma\gamma}(E) \cong 200$ Mpc, 125 Mpc, and 70 Mpc at $E = 1, 3,$ and 10 TeV, respectively,
and a low EBL model based on galaxy counts  \citep{kd10} gives 
 $\lambda_{\gamma\gamma}(E) \cong 280$ Mpc, 150 Mpc, and 85 Mpc, respectively. For analytic
estimates, we write $\lambda_{\gamma\gamma} = 100\lambda_{100}$ Mpc, though
 we use the accurate energy dependence of $\lambda_{\gamma\gamma}(E_{\rm TeV})$
in the numerical calculations. 
The importance of pair-cascade radiation with angular
extent broader than the Fermi LAT psf
depends on the value of 
\begin{equation}
{\lambda_{psf}\over \lambda_{\gamma\gamma}} \cong {{d\theta_{psf}(E_{\rm GeV})
/ \theta_{dfl}}\over
\lambda_{\gamma\gamma}} \cong {\tau_{\gamma\gamma}(E_{\rm TeV})
\theta_{psf}(E_{\rm GeV})
\over \theta_{dfl}}\;,
\end{equation}
where $\lambda_{psf}$ is the effective distance a primary photon would have to 
travel to make a GeV photon detected at the edge of the Fermi psf given the 
parameters of the IGM.
{ The value of $\theta_{psf}(E_{\rm GeV})$,
taken here as the 95\% Fermi LAT
confinement angle, is from the Fermi LAT instrument performance page\footnote{www-glast.slac.stanford.edu/software/IS/glast$\_$lat$\_$performance.htm  } 
\citep[see also][]{ran09,tvn11}.
For the EBL model of Finke et al. (2010), the cascade emission can be treated
as a point source when $B_{-15} \ll 0.05 E_{\rm GeV}^{0.6}$ 
for $0.2 \lesssim E_{\rm GeV} \lesssim 20$.}

For a source at distance $d= d_{\rm Gpc}$ Gpc, with $d_{\rm Gpc} \sim 1$ corresponding to $z \sim 0.2$,  the time delay for emission observed at angle 
\begin{equation}
\theta \cong 0.01\;{\lambda_{100}\over d_{\rm Gpc}}\;\big({B_{-15} w\over E/{\rm 10~GeV}}\big)\;
\label{theatnew}
\end{equation}
 from the line of sight is given from equation (\ref{Deltat}) by 
\begin{equation}
\Delta t ({\rm yr}) \cong 2\times 10^6\;\lambda_{100} \big({B_{-15} w\over E/{\rm 10~GeV}}\big)^2\;
\label{Deltat1}
\end{equation}
Short delay times are restricted to conditions of small $B_{\rm IGMF}$ and large $E$ where, as just seen,
extended pair halo emission can be neglected. 

Equation (\ref{Deltat1}) shows that small time delays are implied when  $\lambda_{\gamma\gamma}$ is small and $\lambda_{psf}/\lambda_{\gamma\gamma}> 1$.
 When $\lambda_{\gamma\gamma}\lesssim \lambda_{\rm T}$, an additional delay $\approx \lambda_{\rm T}\theta_{\rm dfl}^2/c$ arises  during the time that the electrons are losing energy and being deflected by the IGMF \citep[][]{mur08,iit08,rmz04}. Such small values of $\lambda_{\gamma\gamma}\sim 1$ Mpc are only relevant at low redshifts for $\gtrsim 100$ TeV photons that pair-produce within $\approx 1$ Mpc of their source, where the magnetic field may not be representative of the dominant volume of the voids. 
%Thus we can dismiss such an origin of a short time delay without assuming special properties of the TeV sources. The case $\lambda_{\gamma\gamma}\sim d$, %corresponding to the case when the mean free path for $\gamma\gamma$ pair production is about equal to the source distance, formally gives short time delays. 
%This  occurs for $\gamma$ rays with energies of several hundred GeV when $d \sim 1$ Gpc. 
%The  scattered CMB photons hardly reach $ 100$ MeV, and the 
%attenuation of the high-energy photons takes place over the entire distance $d$, so that the fraction of
%upscattered photons with short delay time comprise only a very small fraction of the total.  
%A more robust and realistic alternative to assuming that TeV blazars are steady on timescales of millions of years is to 
%only make claims about the strength of the IGMF from the available TeV data.

\section{$\gamma$ Ray Data of 1ES 0229+200}

%In terms of $\Delta L/\Delta t$, where $\Delta L$ is the change in apparent isotropic luminosity taking place during observing time $\Delta t$, TeV blazars %are, next to GRBs, the most highly variable sources in nature. Indeed, TeV blazars are in some respects more remarkable than GRBs. The TeV blazars PKS 2155-%304 \citep{aha07a} and Mrk 501 \citep{alb07} vary on timescales of a few minutes, shorter than the light crossing time defined by the Schwarzschild radius %of a black hole with mass $\sim 10^9 M_\odot$ that is inferred to power these blazars. The source 1ES 1101-232 shows X-ray flux variations over periods of %several days \citep{aha07b}, and an associated SSC component could lead to TeV flux variations, even though TeV variability has not yet been observed from %this source.

The TeV blazar 1ES 0229+200, which provides some of  the strongest constraints on the lower limit to the IGMF field strength, was observed with HESS \citep{aha07} in 2005 and 2006 
%1 Sept 2005 + 36 d; 20 Agu 2006 + 121 d
and with VERITAS \citep{per10} in October 2009 -- January 2010. No evidence for variability of the TeV flux has been reported, so the observations give an average TeV flux from this source on timescales of $\approx 3$ yr, 
though with poor sampling. The HESS { and preliminary VERITAS data \citep{per10}} are shown in Fig.\ 2 by the blue open circles and red squares, respectively.

Fermi Large Area Telescope upper limits on TeV blazars were reported 
previously \citep{abd09,abd10}. Here we reanalyze 
the Fermi LAT data for 1ES 0229+200 collected from 2008 
August 4 to 2010 September 5 in survey mode. To  minimize systematics, 
only photons with energies greater than 100 MeV were  considered in 
this analysis. In order to avoid contamination from Earth-limb  
$\gamma$ rays, a selection on events with zenith angle $<105^{\circ}$ was 
applied  \citep{atw09}.   This analysis was performed using the standard likelihood analysis
   tools that are part of the Fermi ScienceTools software package
   (version v9r15p5).\footnote{http://fermi.gsfc.nasa.gov/ssc/.} The P6\_V3\_DIFFUSE  set of  instrument response 
functions was used. Photons were selected in a circular region of 
interest (ROI) 10$^\circ$ in radius, centered at the position of 1ES 
0229+200.  The isotropic background, including the sum of  residual 
instrumental background and extragalactic diffuse $\gamma$-ray 
background, was modeled by fitting this component at high galactic 
latitude (isotropic$\_$iem$\_$v02.txt,  available from the FSSC website).  The Galactic diffuse 
emission model version ``gll\_iem\_v02.fit," was used in the analysis. 
The profile likelihood method \citep{rol05} was used to extract 95\% confidence level upper 
limits at the location of 1ES0229+200 assuming a power-law energy distribution with 
photon index=2,   all 1FGL point sources lying within the ROI being 
modeled with power-law distributions. The upper limits shown in Figure 2 
are obtained in the energy bins 0.1 -- 1 GeV, 1 -- 3 GeV, 3 -- 10 GeV, 
1 -- 10 GeV, and 10 -- 100 GeV.

\begin{figure}[t]
% \vspace*{-2.0 cm}
\begin{center}
 \includegraphics[width=2.5in]{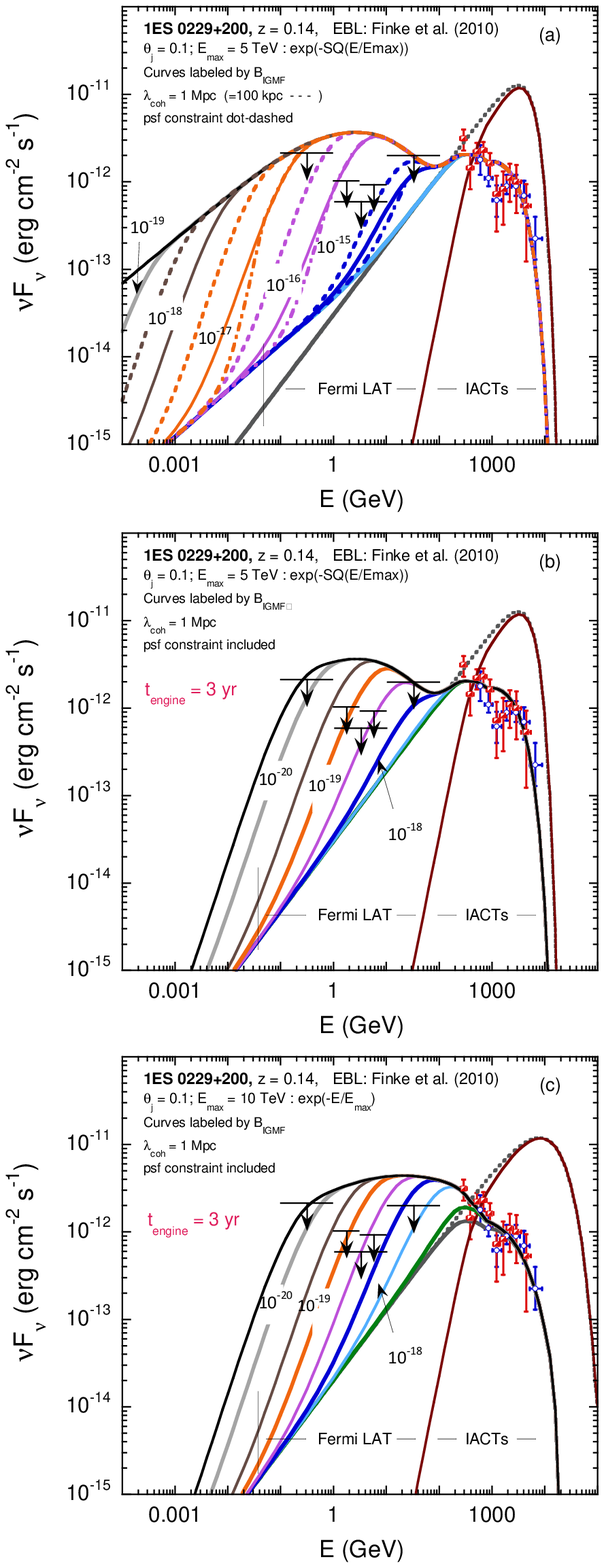} 
% \vspace*{-1.0 cm}
 \caption{Model of cascade radiation spectrum, equation (\ref{fes}), 
applied to HESS,
 VERITAS, 
and Fermi observations of 1ES 0229+200, using model spectra (solid curves) and
 EBL model of \citet{frd10} 
  to give attenuated source spectrum (dotted curves).  (a) Cascade spectra for 
1ES 0229+200 assuming persistent TeV emission at the level observed with HESS and VERITAS, for
different values of $B_{\rm IGMF}$ and $\lambda_{coh} = 1$ Mpc (solid) or $\lambda_{coh} = 100$ kp (dot-dashed). 
The psf constraint for the  $\lambda_{coh} = 1$ Mpc case is shown by the dashed curves.
Cascade spectra when source radiates TeV flux for 3 yr with constant average spectrum
given by power-law with $\nu F_\nu$ index $=4/5$ for source spectrum with superexponential cutoff
$\propto \exp[-(E/5{\rm~TeV})^2]$ (b)  and exponential cutoff
$\propto \exp(-(E/10{\rm~TeV})$ (c), are shown for the case $\lambda_{coh} = 1$ Mpc
with different values of $B_{\rm IGMF}$, as labeled. }
\label{fig1}
\end{center}
\end{figure}

\section{Model for Cascade Radiation}

The limits on the IGMF can be established by employing a simple semi-analytic model for the 
cascade radiation spectrum. Using the notation that $f_\epsilon = \nu F_\nu$ 
at dimensionless photon energy $\epsilon$, and that each photon is attenuated
into a pair with each electron taking one-half the original photon's energy, 
then a straightforward derivation gives
$$f_{\epsilon_s} = {3\over 2} \big({\epsilon_s\over \epsilon_0}\big)^2 
\int_{\max[{\sqrt{\epsilon_s/4\epsilon_0},\gamma_{\rm dfl},\gamma(\Delta t_{eng})}]}^\infty d\gamma
\;\gamma^{-4} \big(1 - {\epsilon_s\over 4\gamma^2 \epsilon_0}\big)\times$$
\begin{equation}
\int_\gamma^\infty d\gamma_i\; 
{f_\epsilon \{\exp[\tau_{\gamma\gamma} (\epsilon,z)]-1\}\over \epsilon^2}\;,
\label{fes}
\end{equation}
where $\gamma_i = \epsilon/2$. The interior integrand represents the fraction of 
deabsorbed source photon flux converted to pairs, and the exterior integral 
represents the Compton-scattered spectrum from cooled electrons \citep[cf.][]{rmz04,mur08,iit08}.
The opacity due to EBL attenuation for photons with measured dimensionless energy
$\epsilon$ from a source at redshift $z \ll 1$ is $\tau_{\gamma\gamma}(\epsilon,z)$, 
and depends on the EBL model.

Equation (\ref{fes}) employs the isotropic Thomson kernel, with the CMB radiation approximated as a monochromatic
radiation field, { but the results in Figure 2 are also integrated over the 
energy distribution of the blackbody radiation field.} The use of the KN kernel makes negligible difference
for photons with energy $\lesssim 20$ TeV. In the three terms in the lower limit of the exterior integration, 
 the first gives the kinematic minimum electron Lorentz factor
to scatter a CMB photon to energy $\epsilon_s$. The second is the value of the deflection 
Lorentz factor $ \gamma_{\rm dfl}$ obtained by equating the Thomson cooling time and the timescale
$\theta_{\rm j} r_{\rm L}/c$
when the electron is deflected outside the photon beam of opening angle $\theta_{\rm j}$. 
{ The third limit, $\gamma(\Delta t_{eng})$, represents the Lorentz factor to which electrons have
cooled after the blazar engine has been operating for time $\Delta t_{eng}$, and follows 
from equation (\ref{Deltat}) by  solving $\Delta t(\gamma_{eng}) < \Delta t_{eng}$ for $\gamma_{eng} = \gamma(\Delta t_{eng})$.
 Here we approximate  $\lambda_{\gamma\gamma}(E_{\rm TeV})  \approx d/\tau_{\gamma\gamma}(E_{\rm TeV})$ Mpc,
using a fit to the Finke et al. (2010) EBL model for 1ES 0229+200. A calculation with 
$\lambda_{\gamma\gamma}( E_{\rm TeV})  \approx d/(2 \tau_{\gamma\gamma}(E_{\rm TeV}))$ Mpc
gives similar results.
Only the first generation of cascade emission attenuated by the factor $\exp[-\tau_{\gamma\gamma}(\epsilon_1,z)]$  is 
shown here.

Results of calculations using this simplified analytic model are shown in Figure 2.
Fig.\ 2a is a calculation where the blazar engine operates for
indefinitely long times, with the reduction of cascade flux due to deflection away from the beam for 
a jet and the detection of a plateau flux of isotropized radiation determined by the
jet opening angle $\theta_{\rm j} =0.1$ \citep{tav10a}.
The source spectrum is described by a super-exponential cutoff power law $\nu F_\nu \propto E^{4/5}\exp[-(E/5{\rm~TeV})^2]$
in Figs. 2(a) and 2(b), and by an exponential cutoff power law $\nu F_\nu \propto E^{4/5}\exp(-E/10{\rm~TeV})$
in Fig.\ 2(c).} 
In agreement
with previous results \citep{nv10,tav10a,tav10b,dol10}, a value of $B_{\rm IGMF}\gtrsim 3\times 10^{-16}$ G is implied in order
to reduce the GeV flux below the Fermi upper limit. { From the calculations, we also find that
under the assumption of persistent TeV 
blazar emission, halo emission becomes increasingly dominant for large jet opening angles. Detection of halos
around AGNs, as claimed by \citet{ak10} \citep[cf.][]{ner11}, would 
then favor detection in sources with large opening angle, long lived TeV engines.
Also under the persistent emission hypothesis,
a maximum jet opening angle $\theta_{\rm j} \lesssim 0.4$ is implied in order that the isotropized 
radiation does not violate the Fermi LAT upper limits.

The effects of $B_{\rm IGMF}$ on 
the received spectrum of reprocessed TeV 
radiation when the blazar engine is assumed to emit
a constant TeV flux over an engine time $\Delta t_{eng} \cong 3$ yr are shown in Figs.\ 2a and 2b. 
These calculations show that $B_{\rm IGMF}\gtrsim 3\times 10^{-19}$ G for the case
where the assumed source spectrum is sharply cut off above 5 TeV. Uncertainties in the analytic model, 
including the strong sensitivity of the cascade spectrum on $\gamma_{eng}$, relaxes 
our conclusions to an analytic, order-of-magnitude minimum IGMF of $B_{\rm IGMF}\gtrsim 10^{-18}$ G for  $\Delta t_{eng} \cong 3$ yr.
Fig.\ 2c shows that the minimum magnetic field also depends sensitively on the characterization 
of the high-energy spectral flux, which can then quickly cascade into the 10 -- 100 GeV band and violate a Fermi upper 
limit (or detection; see \citet{okd11}). By assuming source spectra with larger fluxes above $\approx 5$ --  10 TeV, \citet{dol10} and \cite{tvn11} derive
larger values for the minimum $B_{\rm IGMF}$, but not more than a factor of a few above the analytic results when difference in 
activity times and primary source fluxes are considered.}

\section{Discussion and Summary}

Previous GeV/TeV inferences of the strength of the intergalactic magnetic field make an assumption 
that the mean blazar TeV flux over millions of years remains similar to values observed over the last few few years.
Our knowledge of the blazar engine is not { yet} so good as to have high confidence in this assumption, 
though some models for slowly varying TeV flux from TeV blazars can be noted. 
For example, a slow cooling rate of the electrons that make the TeV photons 
could imply a slowly varying $\gamma$-ray flux even if the blazar engine is very active.
 For electrons scattering photons to TeV energies, 
the synchrotron cooling time for the observer is $t_{syn} \cong (1+z) 6\pi m_ec/(\delta_{\rm D} \sigma_{\rm T} B^{\prime 2} \gamma^\prime) \cong
50/E({\rm TeV})$ yr, using the fitting parameters of \citet{tav10c} for 1ES 0229+200 
(break Lorentz factor $\gamma_{br} = 5\times 10^6$, emission region magnetic field $B^\prime = 5\times 10^{-4}$ G, and Doppler factor $\delta_{\rm D} = 40$). 
 Relativistic electrons in an extended jet that Compton scatter photons of the CMB could also make slowly varying TeV radiation in  sources like 1ES 0229+200 or 1ES 1101-232 \citep{bdf08}. In this model, relativistic electrons  lose energy on timescales of $\approx 750/[(\Gamma/10)^2\sqrt{E({\rm TeV})}]$ yr.  
These models do not, however, provide good reasons to expect TeV blazars to produce steady flux for thousands or millions of years. 

A more reliable limit
is obtained from direct measurements of TeV fluxes. For
the handful of observations of 1ES 0229+200 over 3 -- 4 years of observing \citep{aha07,per10},
no TeV flux variations have been reported. Using such timescales leads to a 
limit of $B_{\rm IGMF}$(G) $\gtrsim
10^{-18}(E/10{\rm ~GeV})\sqrt{\Delta t/3{\rm ~yr}}/\sqrt{\lambda_{100}}$, 
assuming that $\lambda _{coh} \approx 1$ Mpc. { By assuming strong intrinsic $\gtrsim 10$ TeV emission 
from 1ES 0229+200 (which is not observed because 
of EBL attenuation),  Fermi LAT flux upper limits at $\approx 100$ GeV can be violated, leading to larger 
limiting values of 
 $B_{\rm IGMF}$(G) $\gtrsim 5\times 10^{-18}$ G. { Evidence for a strong primary flux at $\gtrsim 10$ TeV comes 
from detection of a shoulder feature at $\approx 1$ TeV, as found in the numerical calculations \citep{dol10} and
analytical results (Fig.\ 2c), and suggested by the joint VERITAS/HESS data.} 
Note that our calculations assume negligible contribution from cascades induced by
photopair interactions by $\gtrsim 10^{18}$ eV cosmic rays \citep{ess10}.
More frequent, sensitive, and broadband GeV -- TeV observations of 1ES 0229+200
can test whether the average TeV flux corresponds to the flux that has  been historically measured or is unusual.}

Evidence for long-lived TeV radiation can be identified in pair halos \citep{acv94} from 
misaligned blazar candidates such as Cen A or M87. Searches for pair echoes from GRBs, which are sensitive at $\ll
10^{-21}/\lambda_{coh}({\rm Mpc})$ G \citep{tak08}, would test our claim that $B_{\rm IGMF}\gtrsim 10^{-18}$ G.
A large field-of-view detector like the High Altitude Water Cherenkov telescope \citep{goo10}, or systematic monitoring 
campaigns of blazars like 1ES 0229+200, 1ES 1101-232 ($z = 0.186$), 1ES 0347-121 ($z = 0.185$) or other bright, moderate redshift BL Lacs with 
the present generation of air Cherenkov telescopes or an advanced Cherenkov telescope array, will give better information about 
the duty cycle of TeV blazars and provide more secure constraints on the value of the 
intergalactic magnetic field.

\acknowledgements 
{ We thank J. Perkins for discussions about the VERITAS data, the VERITAS team for 
kindly allowing us to show their preliminary 1ES 0229+200 data, and 
the referee for a constructive report.}
This work is supported by NASA Fermi Guest Investigator Program DPR 76-644-10. M.C.\ 
acknowledges the support and hospitality of the NRL
High Energy Space Environment Branch during his
visit. The work of C.D.D.\ and J.D.F. is also supported by the
Office of Naval Research. 

The $Fermi$ LAT Collaboration acknowledges support from a number of agencies and institutes for both development and the operation of the LAT as well as scientific data analysis. These include NASA and DOE in the United States, CEA/Irfu and IN2P3/CNRS in France, ASI and INFN in Italy, MEXT, KEK, and JAXA in Japan, and the K.~A.~Wallenberg Foundation, the Swedish Research Council and the National Space Board in Sweden. Additional support from INAF in Italy and CNES in France for science analysis during the operations phase is also gratefully acknowledged.

\end{document}